\documentclass[11pt]{article}
\usepackage[margin=1in]{geometry}

\usepackage{microtype}
\usepackage{amsmath}
\usepackage{booktabs}
\usepackage{tabularx}
\usepackage{enumitem}
\usepackage{xcolor}
\usepackage{listings}
\usepackage{tikz}
\usetikzlibrary{arrows.meta,positioning,calc,fit}
\usepackage[hidelinks]{hyperref}
\usepackage{url}

\definecolor{accent}{HTML}{2F6FD6}
\newcolumntype{Y}{>{\raggedright\arraybackslash}X}
\newcolumntype{L}[1]{>{\raggedright\arraybackslash}p{#1}}
\setlist{itemsep=2pt,topsep=4pt}

\title{Agent Authenticator: Passkey-Grade, Hardware-Bound\\Authentication for AI Agents}
\author{Harish Santhanalakshmi Ganesan\\
\small Independent Researcher\\
\small\texttt{harishsg99@gmail.com}}
\date{November 2026}

\begin{document}
\maketitle

\begin{abstract}
AI coding agents now push code, publish packages, call model APIs and operate cloud resources on behalf of developers. They authenticate to these services with the same bearer secrets humans use, personal access tokens, API keys and package-registry tokens, stored where the model, its subprocesses and any co-resident malware can read them. Incidents in 2025 and 2026, including prompt-injection attacks on coding agents and self-propagating supply-chain worms, show that this arrangement fails in predictable ways. Passkeys solved the analogous problem for humans, but they require a human gesture on every use and cannot be operated by an autonomous process. We present the \emph{Agent Authenticator}, a design that gives agents passkey-grade authentication properties: per-service hardware-bound keys, no shared secrets, origin binding enforced by the runtime rather than the model, normalized attestation of the execution environment, human-signed scoped mandates, and cross-device human approval for high-risk actions. The same protocol serves agents on developer laptops and in cloud enclaves, and it deploys as a profile over existing mechanisms, MCP authorization, OAuth token exchange including the Identity Assertion JWT Authorization Grant, git and SSH, rather than as a new identity system. The authenticator also produces a signed, hash-chained record of each action it authorizes. We analyze the design against a threat model centred on prompt injection and same-host compromise, map it onto five documented incidents, and discuss its limits.
\end{abstract}

\section{Introduction}

Software agents built on large language models have moved from suggesting code to acting on it. A coding agent today may read a repository, run tests, push a branch, open a pull request, comment on an issue tracker through a Model Context Protocol (MCP) server, publish a package, or provision cloud resources. Each of these actions requires authenticating to a service that was designed for human users or for traditional workloads.

In practice, agents inherit whatever credentials sit in the developer's environment: an SSH key, a GitHub personal access token, a cloud CLI profile, an npm token in a configuration file. These are bearer secrets. Whoever holds them can use them, from any machine, for anything they permit. Three properties of agents make this arrangement worse than it is for human developers. First, agents read untrusted content as part of their work (issues, pull-request titles, README files, dependency source), and that content can contain instructions; prompt injection turns any text the agent reads into a potential command. Second, agents spawn subprocesses whose behaviour the agent's own safeguards do not observe. Third, agents act at machine speed without a human checking each step.

Section~\ref{sec:problem} documents five incidents from 2025 and 2026 in which these properties led to credential theft or misuse, and reviews data showing that static secrets remain the norm in agent tooling. The pattern is consistent: credentials were ambient, broadly scoped, exportable, and protected only by controls running inside the agent.

For human users, the industry's answer to phishing and credential theft has been passkeys~\cite{webauthn3,ctap22}: per-site key pairs held in secure hardware, with the browser rather than the user deciding which site a key may be used for. Passkeys cannot be used by agents directly. They require a human gesture for each assertion and offer no programmatic enrollment. Yet their core properties are exactly what agents lack.

This paper asks what a passkey for agents should look like, and answers with a design we call the \emph{Agent Authenticator}. Our contributions are:

\begin{enumerate}
\item An analysis of documented agent-credential incidents that identifies four recurring failure classes (Section~\ref{sec:problem}).
\item A design that transfers passkey properties to autonomous agents and adds three properties passkeys lack: scoped authority through human-signed mandates, accountability of every assertion to a named human, and identity-bound action records (Sections~\ref{sec:design} and~\ref{sec:records}).
\item A deployment model showing how the design layers onto MCP authorization, OAuth token exchange, the Identity Assertion JWT Authorization Grant (ID-JAG), git, SSH and cloud workload identity, with one protocol for local and cloud agents (Section~\ref{sec:deploy}).
\item A security analysis against an explicit threat model, including a mapping of the five incidents onto the design and a statement of residual risk (Section~\ref{sec:analysis}).
\end{enumerate}

This paper presents the design rationale and analysis; we have not yet built a full implementation, and we state the resulting limitations in Section~\ref{sec:limits}.

\section{Background}
\label{sec:background}

\paragraph{Passkeys.} WebAuthn~\cite{webauthn3} and CTAP~\cite{ctap22} define public-key credentials scoped to a relying party (RP) identifier. The authenticator holds the private key, often in secure hardware, and signs a challenge together with authenticator data that includes flags for user presence (UP) and user verification (UV) and a signature counter. The browser, not the user, supplies the origin, which is what makes passkeys phishing-resistant: a credential created for one site cannot be exercised by another.

\paragraph{OAuth delegation.} OAuth 2.0 Token Exchange~\cite{rfc8693} lets a client exchange one security token for another, carrying both a \emph{subject} (on whose behalf) and an \emph{actor} (who is acting). The JWT authorization grant profile~\cite{rfc7523} lets a signed assertion be redeemed for an access token. Rich Authorization Requests~\cite{rfc9396} express fine-grained permissions as structured \texttt{authorization\_details}. DPoP~\cite{rfc9449} binds access tokens to a client-held key so that a stolen token is useless without that key. The Identity Assertion JWT Authorization Grant~\cite{idjag} combines token exchange and JWT grants so that an enterprise identity provider (IdP) can broker an application's access to another application; enterprise IdPs have adopted it for agent access, and MCP uses it for enterprise-managed authorization~\cite{mcpauth}.

\paragraph{Remote attestation.} The RATS architecture~\cite{rfc9334} separates an \emph{attester} that produces evidence about its state from a \emph{verifier} that appraises it. The Entity Attestation Token (EAT)~\cite{rfc9711} is a standard claims format for such evidence. Platform mechanisms that can attest to hardware key storage include TPM 2.0 key attestation, Apple App Attest, Android key attestation, and confidential-computing reports from AMD SEV-SNP, Intel TDX and AWS Nitro Enclaves.

\paragraph{Transparency logs.} SCITT~\cite{rfc9943} defines transparency services that register signed statements and return receipts, making later deletion or equivocation detectable.

\section{Problem Analysis}
\label{sec:problem}

\subsection{Incidents}

Table~\ref{tab:incidents} summarizes five documented incidents. We selected incidents that involved coding agents or developer credentials, were publicly analyzed, and illustrate distinct failure mechanisms.

\begin{table}[t]
\centering\small
\caption{Documented incidents involving agent or developer credentials.}
\label{tab:incidents}
\begin{tabularx}{\linewidth}{@{}L{2.6cm}YY@{}}
\toprule
Incident & What happened & Root cause \\
\midrule
GitHub MCP prompt injection (2025)~\cite{invariant} & A malicious issue in a public repository caused an agent to read private repositories and publish their contents in a public pull request. & One broadly scoped token covered every repository the user could reach. \\
Comment and Control (2026)~\cite{csa} & A crafted pull-request title hijacked several CI-hosted coding agents, which leaked repository secrets back through the code host itself. & Secrets were ambient in the agent's environment; untrusted text was treated as instructions. \\
Coding-agent exploits (2025--2026)~\cite{venturebeat} & Crafted branch names exfiltrated tokens; repository settings files bypassed trust prompts; write sandboxes were escaped. & Controls ran inside the agent and could be disabled by repository content. \\
Nx ``s1ngularity'' (2025)~\cite{nx} & Malicious package versions harvested 2,349 credentials from 1,079 developer machines, invoking locally installed AI CLIs to search for secrets. & Tokens and keys on disk were readable by any local process. \\
Shai-Hulud worm (2025--2026)~\cite{shaihulud} & A self-propagating worm stole npm and GitHub tokens and republished itself into packages the victims maintained. & Long-lived publish tokens were usable from any machine. \\
\bottomrule
\end{tabularx}
\end{table}

These incidents are consistent with broader measurements. An analysis of 5,205 open-source MCP servers found that 88\% require credentials, 53\% rely on static API keys or personal access tokens, and only 8.5\% use OAuth~\cite{astrix}. A 2026 survey of public GitHub found 28.65 million newly leaked secrets in 2025, an 81\% rise in leaked AI-service secrets, and 24,008 secrets in MCP configuration files, of which 2,117 were valid~\cite{gitguardian2026}.

\subsection{Failure classes}

Across the incidents we identify four failure classes:

\begin{description}[leftmargin=1em]
\item[F1 Ambient bearer secrets.] Credentials are readable by the agent, its subprocesses and co-resident code, and usable from anywhere.
\item[F2 In-agent controls.] Safeguards run inside the process they are meant to constrain, so content that subverts the agent can also subvert its safeguards.
\item[F3 Over-broad scope.] Credentials authorize far more than the task requires and persist far longer.
\item[F4 Missing attribution.] Services cannot distinguish an agent's actions from its human's, and nothing binds an action to the person who authorized it.
\end{description}

\subsection{Why existing mechanisms fall short}

Passkeys address F1 for humans but cannot be operated autonomously. OAuth 2.1 and MCP authorization address F3 with scopes and expiry but issue bearer tokens by default and say nothing about where the client's keys live. ID-JAG lets an IdP decide which applications an agent may reach, but it is anchored in the user's login session and carries no evidence about the agent's key or execution environment. Workload identity systems remove static secrets for cloud services but have no notion of a human approver and fit developer laptops poorly. Secret vaults and proxies keep secrets out of the model's context, addressing part of F1, but the secrets remain bearer tokens and the approach is not standardized. No existing mechanism addresses all four classes together.

\section{Threat Model and Goals}
\label{sec:threat}

\paragraph{Adversaries.} We consider four adversaries.
\begin{itemize}
\item \textbf{A1, content injector:} controls text the agent reads (issues, pull requests, documentation, tool outputs, dependency source) and can therefore steer the model's outputs and tool calls.
\item \textbf{A2, co-resident code:} malicious code running on the same host as the agent with the same user privileges, for example a compromised dependency's install script.
\item \textbf{A3, network and origin attacker:} operates servers the agent might be induced to contact and can observe or replay traffic it receives.
\item \textbf{A4, compromised cloud image:} controls the software image of a cloud sandbox in which an agent runs, but not the underlying confidential-computing hardware.
\end{itemize}

\paragraph{Out of scope.} An adversary with administrator or kernel privileges on a developer machine, physical attacks on secure hardware, a compromised IdP, and a human approver who knowingly approves harmful actions. We discuss the consequences of these exclusions in Section~\ref{sec:limits}.

\paragraph{Goals.}
\begin{description}[leftmargin=1em]
\item[G1] No credential usable to mint tokens ever reaches the model process.
\item[G2] A1 cannot cause an agent to authenticate to an origin for which no credential was provisioned.
\item[G3] Agent keys are non-exportable, and relying parties can learn, in a normalized form, how strongly they are protected.
\item[G4] Every credential and every high-risk action is approved by a named human, possibly on a different device from the agent.
\item[G5] Authority is scoped to services, actions, resources and time, and can only narrow under delegation.
\item[G6] The design deploys over existing IdPs and services.
\item[G7] One protocol serves local and cloud agents.
\end{description}

\section{Design}
\label{sec:design}

\subsection{Architecture}

\begin{figure}[t]
\centering
\begin{tikzpicture}[
  box/.style={draw,rounded corners=3pt,align=center,minimum width=3.2cm,minimum height=1.15cm,font=\small},
  main/.style={box,draw=accent,very thick,fill=accent!8},
  arr/.style={-{Stealth[length=2.2mm]},thick},
  lbl/.style={font=\footnotesize,fill=white,inner sep=1.5pt}]
\node[box] (agent) at (0,0) {\textbf{Agent}\\model, tools\\holds no keys};
\node[box] (rt) at (5.2,0) {\textbf{Agent runtime}\\sets origin from\\live TLS};
\node[box] (rp) at (10.4,0) {\textbf{Relying parties}\\code host, registry,\\MCP, cloud};
\node[box] (idp) at (10.4,2.6) {\textbf{Enterprise IdP}\\policy, revocation};
\node[box] (human) at (0,-2.8) {\textbf{Human approver}\\passkey on phone\\or laptop};
\node[main] (aa) at (5.2,-2.8) {\textbf{Agent Authenticator}\\keys in TPM / enclave\\checks mandate, signs};
\draw[arr] (agent) -- node[lbl,above=2pt]{request} (rt);
\draw[arr] (rt) -- node[lbl,above=2pt]{assertion} (rp);
\draw[arr] ([xshift=-5mm]rt.south) -- node[lbl,left=2pt]{sign request} ([xshift=-5mm]aa.north);
\draw[arr] ([xshift=5mm]aa.north) -- node[lbl,right=2pt]{signed assertion} ([xshift=5mm]rt.south);
\draw[arr] (human) -- node[lbl,above=2pt]{approval} (aa);
\draw[{Stealth[length=2.2mm]}-{Stealth[length=2.2mm]},thick] (rp) -- node[lbl,left=2pt]{token exchange} (idp);
\draw[arr] (idp.east) -- ++(0.6,0) |- node[lbl,pos=0.75,below=2pt]{revocation signals} (aa.east);
\end{tikzpicture}
\caption{Roles in the Agent Authenticator design. Only the authenticator holds keys; the runtime fixes the origin; the IdP keeps policy authority.}
\label{fig:arch}
\end{figure}
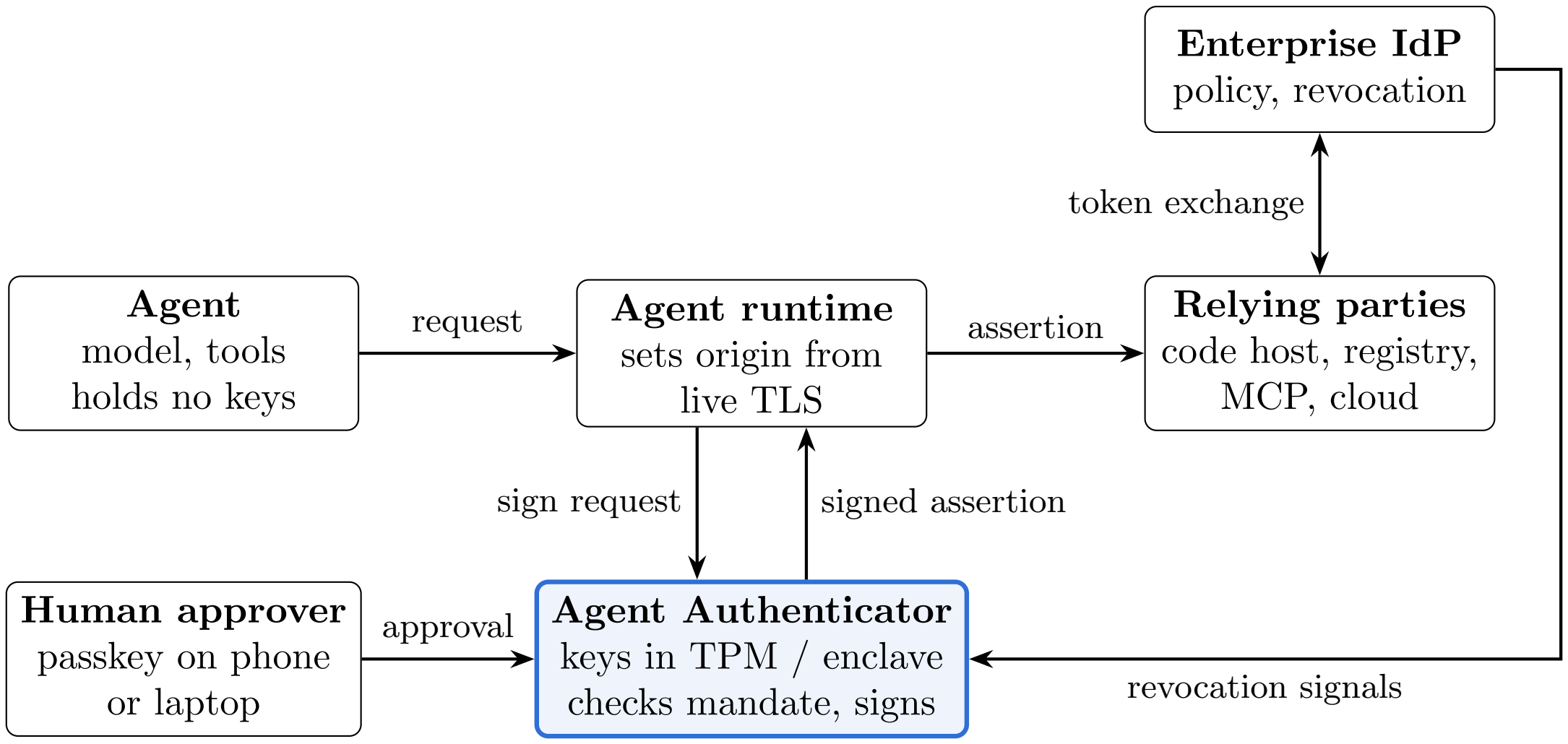

Figure~\ref{fig:arch} shows the roles. The \emph{agent} is the model-driven process and everything it spawns; we treat it as untrusted. The \emph{agent runtime} is the harness that launches the agent and owns its network connections. The \emph{Agent Authenticator} (AA) is a daemon running in a separate privilege domain that holds keys in secure hardware, enforces mandates and signs. \emph{Relying parties} are the services the agent calls. An enterprise \emph{IdP} may sit in front of relying parties as the policy authority. The \emph{human approver} holds an ordinary passkey.

The central design decision mirrors the browser's role in WebAuthn: the runtime, not the model, supplies the origin of every request, taking it from the live TLS connection. The model can ask for an action but cannot name the origin that the resulting signature is bound to.

\subsection{Keys and assurance levels}

Four keys participate (Table~\ref{tab:keys}); none is accessible to the model process. Agent credential keys are scoped to a single relying party, so a credential for one service is cryptographically useless at any other.

\begin{table}[t]
\centering\small
\caption{Keys in the design.}
\label{tab:keys}
\begin{tabularx}{\linewidth}{@{}lYY@{}}
\toprule
Key & Holder & Lifetime \\
\midrule
Approver key & Human's passkey authenticator & Long-lived; device-bound passkeys recommended \\
AA identity key & AA secure hardware & Life of the device or cloud environment \\
Agent credential key & AA secure hardware, one per agent and relying party & Local: until mandate expiry or revocation. Cloud: one task \\
Evidence key & AA secure hardware, separate from credential keys & Rotated per epoch, forward-secure \\
\bottomrule
\end{tabularx}
\end{table}

Relying parties should not need to understand every vendor's attestation format. The AA converts platform evidence into an EAT~\cite{rfc9711} carrying a normalized \emph{assurance level} (AL), appraised by a RATS verifier~\cite{rfc9334}:

\begin{itemize}
\item \textbf{AL3:} keys in a trusted execution environment that also isolates the agent's code, with the runtime measured (SEV-SNP, TDX, Nitro, Arm CCA).
\item \textbf{AL2:} keys in dedicated secure hardware, host software not isolated (TPM 2.0, Secure Enclave via App Attest, Android StrongBox).
\item \textbf{AL1:} keys in a platform keystore without hardware isolation, or cloud workload identity only.
\item \textbf{AL0:} software keys.
\end{itemize}

Policies are written against levels, for example ``publishing requires AL2 and human approval'', rather than against platforms.

\subsection{Mandates}

A \emph{mandate} is a statement signed by the human approver that bounds what an agent credential may do. It names the agent credential by key thumbprint~\cite{rfc7638}, the AA and its minimum assurance level, the permitted relying parties, the permitted actions expressed as RAR \texttt{authorization\_details}~\cite{rfc9396}, the actions that require fresh human approval, whether sub-delegation is allowed, and an expiry. The approver signs it with an ordinary WebAuthn assertion whose challenge is the mandate's hash, so no change to authenticators is needed.

\begin{lstlisting}
{ "typ": "agent-mandate+jwt", "sub": "<human id>",
  "agent_cred": { "jkt": "<agent key thumbprint>" },
  "aa": { "id": "<AA key thumbprint>", "min_al": "AL2" },
  "rps": ["github.com", "registry.npmjs.org"],
  "authorization_details": [
    { "type": "git", "actions": ["push"],
      "locations": ["github.com/acme/api"], "refs": ["agent/fix-*"] } ],
  "step_up": ["merge", "publish", "delete"],
  "delegable": false, "exp": 1762480000 }
\end{lstlisting}

\subsection{AgentData and assertions}

Every assertion signs an \emph{AgentData} structure, the analogue of WebAuthn's authenticator data. It contains the RP identifier and origin (both supplied by the runtime), the RP's challenge, the credential identifier, the mandate hash, the assurance level, a strictly increasing counter, the action being authorized, and two flags that replace UP and UV: \texttt{mb} (the agent is acting within its mandate) and \texttt{ha} (a human approved this specific action). When \texttt{ha} is set, the assertion also carries the approver's WebAuthn assertion over the hash of the AgentData.

\subsection{Ceremonies}

\paragraph{Registration.} The runtime requests registration from the relying party, which returns a nonce. The AA creates a key pair scoped to that RP in secure hardware and produces an attestation bound to the nonce. It then presents a plain-language rendering of a draft mandate to the human, locally or on their phone, and the human signs it with their passkey. The relying party verifies the attestation, the approver's signature and the mandate, and stores the credential public key as a separately revocable credential linked to the human's account.

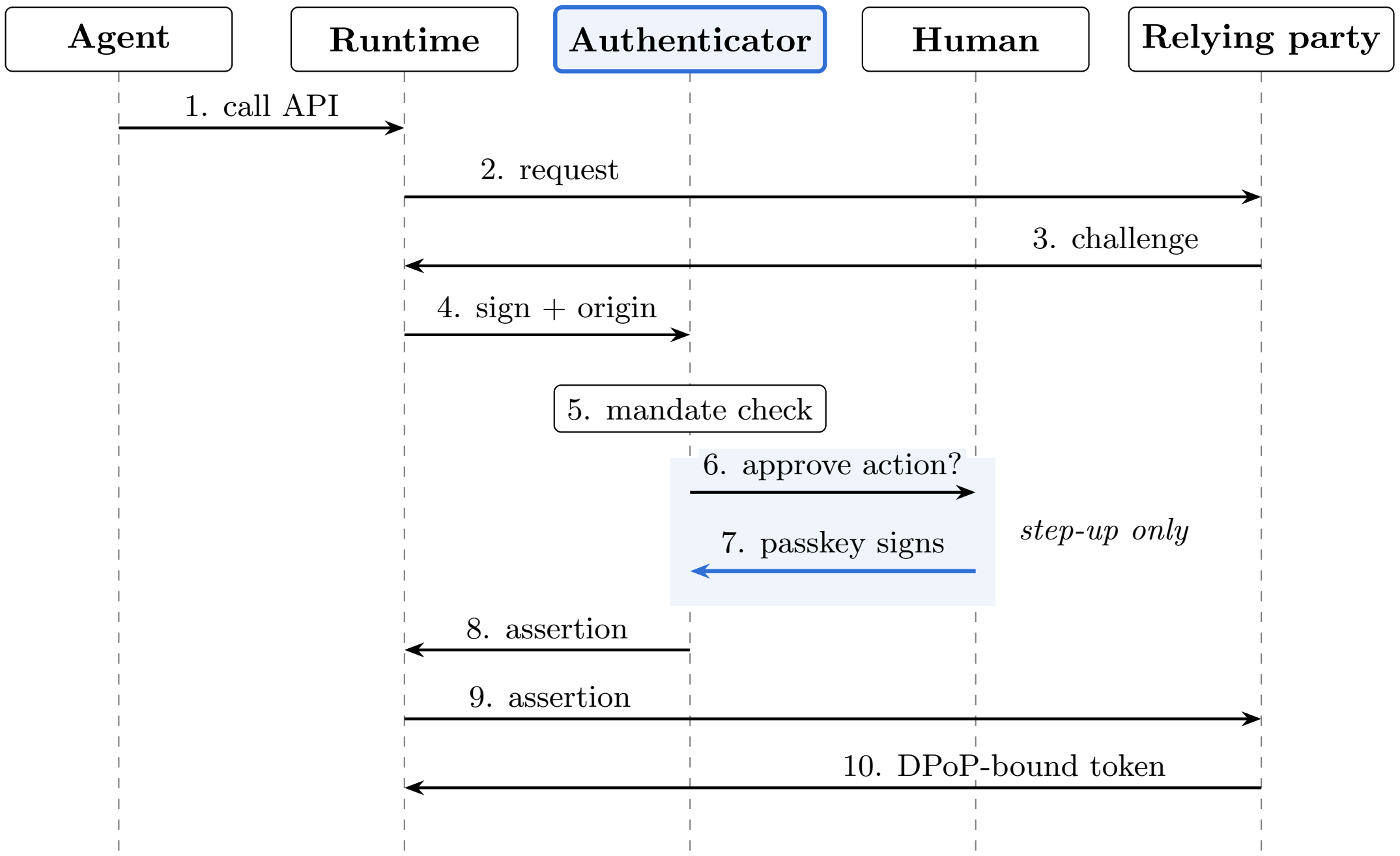
\begin{figure}[t]
\centering
\begin{tikzpicture}[font=\small,
  hdr/.style={draw,rounded corners=2pt,minimum width=2.3cm,minimum height=0.65cm,font=\small\bfseries},
  arr/.style={-{Stealth[length=2mm]},thick},
  lbl/.style={font=\footnotesize,above=2pt,fill=white,inner sep=1pt}]
\foreach \x/\n/\t in {0/ag/Agent,2.9/rt/Runtime,5.8/aa/Authenticator,8.7/hu/Human,11.6/rp/Relying party}{
  \node[hdr] (\n) at (\x,0) {\t};
  \draw[gray,dashed] (\x,-0.33) -- (\x,-8.3);
}
\node[hdr,draw=accent,very thick,fill=accent!8] at (5.8,0) {Authenticator};
\fill[accent!7] (5.6,-4.25) rectangle (8.9,-5.75);
\node[font=\footnotesize\itshape,anchor=west] at (9.0,-5.0) {step-up only};
\draw[arr] (0,-0.9) -- node[lbl]{1. call API} (2.9,-0.9);
\draw[arr] (2.9,-1.6) -- node[lbl,pos=0.17]{2. request} (11.6,-1.6);
\draw[arr] (11.6,-2.3) -- node[lbl,pos=0.17]{3. challenge} (2.9,-2.3);
\draw[arr] (2.9,-3.0) -- node[lbl]{4. sign + origin} (5.8,-3.0);
\node[draw,rounded corners=2pt,fill=white,font=\footnotesize] at (5.8,-3.75) {5. mandate check};
\draw[arr] (5.8,-4.6) -- node[lbl,fill=accent!7]{6. approve action?} (8.7,-4.6);
\draw[arr,accent,very thick] (8.7,-5.4) -- node[lbl,text=black,fill=accent!7]{7. passkey signs} (5.8,-5.4);
\draw[arr] (5.8,-6.2) -- node[lbl]{8. assertion} (2.9,-6.2);
\draw[arr] (2.9,-6.9) -- node[lbl,pos=0.17]{9. assertion} (11.6,-6.9);
\draw[arr] (11.6,-7.6) -- node[lbl,pos=0.3]{10. DPoP-bound token} (2.9,-7.6);
\end{tikzpicture}
\caption{Authentication with optional step-up. Routine calls skip steps 6--7 and are signed on the mandate alone.}
\label{fig:login}
\end{figure}

\paragraph{Authentication and step-up.} Figure~\ref{fig:login} shows authentication. The AA checks the mandate before signing: the RP and action must be permitted, the mandate unexpired, and the assurance level still met. For actions listed as requiring approval, the AA sends a summary of exactly that action to the approver's device; the approval signs the hash of the AgentData, so it covers one action and cannot be replayed. Because the approval travels to the human's own authenticator, the ceremony is identical whether the agent runs on the same laptop or in a remote enclave.

\paragraph{Relying-party verification.} The relying party verifies, in order: the credential exists and is not revoked; the RP identifier and origin match its own; the challenge is fresh and unused; the signature is valid; the counter has increased (a non-increasing counter indicates cloning and triggers revocation); the mandate hash matches the registered mandate, which has not expired; the action is permitted by the mandate; the assurance level meets both the mandate and local policy; and, where required, a valid human approval over this AgentData is attached. Only then does it issue a short-lived access token bound to the agent key with DPoP~\cite{rfc9449}. A relying party may delegate the policy checks to its IdP through token exchange but always performs the cryptographic checks itself.

\paragraph{Sub-delegation.} An agent may hand a narrower task to another agent only if its mandate allows. The child mandate must be a subset of the parent in relying parties, actions and expiry and may not lower the minimum assurance level, so authority narrows monotonically along a delegation chain, and every hop's key must itself be attested.

\paragraph{Revocation.} Mandate expiry, administrator action, IdP session revocation, failed re-attestation or a broken action-record chain revoke a credential. The AA subscribes to OpenID Shared Signals~\cite{ssf} so that a revocation at the IdP destroys the corresponding keys, and it emits signals when attestation fails.

\section{Deployment}
\label{sec:deploy}

\subsection{Bindings to existing protocols}

A design that requires every service to adopt a new protocol will not be adopted. Table~\ref{tab:bindings} lists how the AA plugs into mechanisms services already run. Two points deserve comment. First, ID-JAG requires the user's identity assertion as the token-exchange \emph{subject}; our binding presents the agent assertion as the \emph{actor} token, which matches token exchange's semantics of one party acting for another~\cite{rfc8693} and leaves IdP policy in control. Second, for relying parties that support none of these mechanisms, the AA may obtain legacy tokens through a broker, but it records the action at no higher than AL1 and never exposes the token to the model.

\begin{table}[t]
\centering\small
\caption{Bindings onto existing mechanisms.}
\label{tab:bindings}
\begin{tabularx}{\linewidth}{@{}L{2.6cm}YY@{}}
\toprule
Binding & Mechanism & Role of the AA \\
\midrule
MCP & MCP authorization with DPoP and workload-identity extensions~\cite{mcpauth} & Acts as the MCP client's auth provider; tokens bound to the agent key \\
Enterprise IdP & ID-JAG: token exchange then JWT grant~\cite{idjag,rfc8693,rfc7523} & Agent assertion is the actor token; IdP applies policy \\
OAuth APIs & JWT grant or token exchange with RAR & Assertion redeemed for a key-bound access token \\
Git over HTTPS & git credential helper & Per-remote tokens; no tokens on disk \\
Git over SSH & ssh-agent with a key type modelled on FIDO SSH keys & Signatures carry flags and mandate hash \\
Cloud & Workload identity federation & Short-lived cloud credentials; no static keys \\
Package registries & Trusted publishing & Publishing requires human approval \\
\bottomrule
\end{tabularx}
\end{table}

\subsection{Local and cloud profiles}

The protocol is identical in both settings; only the AA's form and the attainable assurance differ. On a developer machine the AA is a system daemon using the TPM or Secure Enclave, typically reaching AL2. It must run outside the agent's user or sandbox and sign only for processes it launched, identified by process lineage and, where the operating system offers it, code signature. In the cloud, the AA is a sidecar or enclave key service; keys are released only after the environment's attestation matches an expected measurement, reaching AL3 in confidential computing, and credentials last for one task. When a session moves from a laptop to the cloud, the cloud agent receives a fresh credential under a child mandate; the local key is never copied.

\section{Action Records}
\label{sec:records}

Because every credentialed call must pass through the AA to be signed, the AA observes every action that touches a service without monitoring the operating system. It writes one record per action at the moment it signs or refuses it, recording which agent, under which human's mandate, performed which action on which target, with which outcome (allowed, step-up approved or denied, or denied for scope, origin or revocation). Records are signed by the AA's evidence key, never by the agent, and hash-chained; each assertion carries the current chain head, so every relying party holds an independent witness; and chain heads are periodically registered with a SCITT transparency service~\cite{rfc9943}. Local actions that involve no credential, such as file edits, can be reported by the runtime but are labelled as runtime-reported, since the agent's own harness produced them. A signed heartbeat makes an interruption of the AA detectable. These mechanisms are tamper-evident rather than tamper-proof: an administrator on the host can stop the AA, but cannot do so without leaving a provable gap.

\section{Security Analysis}
\label{sec:analysis}

We argue informally that the design meets its goals against the adversaries of Section~\ref{sec:threat}.

\paragraph{G1 against A1 and A2.} The model process never holds a credential: it receives neither private keys, which remain in secure hardware, nor long-lived tokens, and the access tokens issued after authentication are DPoP-bound to the agent key. A1 can at most cause the agent to \emph{request} actions. A2, running with the same user privileges, cannot extract hardware-held keys; it can attempt to ask the AA to sign, which the AA refuses for processes outside the lineage it launched. On platforms where lineage cannot be established reliably this check is weaker; high-risk actions additionally require human approval, which A2 cannot forge.

\paragraph{G2 against A1 and A3.} The origin and RP identifier in AgentData come from the runtime's view of the connection, and credentials are scoped per relying party. An agent induced to contact an attacker's server has no credential for that origin, and an assertion captured by A3 is bound to a different RP identifier, a single-use challenge and a counter, so it cannot be replayed elsewhere. A1 can still misuse relying parties the mandate permits; G5 bounds that damage.

\paragraph{G3 and A4.} Keys are generated in secure hardware and never leave it; the normalized assurance level is derived from appraised attestation evidence. In the cloud profile, keys are released only to environments whose measurements match expectations, so A4's modified image does not receive them.

\paragraph{G4 and G5.} Credentials are unusable until a mandate signed with the approver's passkey is registered, and step-up approvals sign the specific AgentData, binding a human decision to one action. Mandates use RAR's structured permissions, and child mandates must be subsets of their parents.

\paragraph{Mapping to incidents.} Table~\ref{tab:mapping} states, for each incident, the expected outcome under the design. These are qualitative judgments derived from the incident reports and the design, not measurements.

\begin{table}[t]
\centering\small
\caption{Expected outcome of each incident under the design (qualitative).}
\label{tab:mapping}
\begin{tabularx}{\linewidth}{@{}L{3.1cm}L{1.7cm}Y@{}}
\toprule
Incident & Outcome & Mechanism \\
\midrule
GitHub MCP injection & Contained & Mandate limits the agent to named repositories; cross-repository publication outside them is refused. \\
Comment and Control & Blocked & No secrets exist in the agent's environment; review agents receive no secret-reading scope. \\
Coding-agent exploits & Contained & Policy is enforced by the AA outside the agent, so repository content cannot disable it; tokens are not exposed to exfiltrate. \\
Nx s1ngularity & Blocked & No credentials on disk to harvest; hardware keys cannot be read. \\
Shai-Hulud worm & Blocked & Publishing requires per-action human approval, and credentials cannot be used from other machines. \\
\bottomrule
\end{tabularx}
\end{table}

\section{Related Work}
\label{sec:related}

\paragraph{Standards work on agent identity.} The IETF WIMSE working group's AIMS document~\cite{aims} composes workload identity, OAuth and shared signals into a model for agent authentication and authorization, treating agents as workloads; it does not bind agent keys to devices or to a human approver's passkey. AAuth~\cite{aauth} gives each agent its own signing key and signs every request, with optional platform attestation in its bootstrap guidance. The PTV draft~\cite{ptv} binds agent identity to TPM-anchored attestation of model and policy, and condition-bounded credentials~\cite{winmagic} make TPM-resident keys usable only while a verified user is present on an endpoint. That design centres on endpoint-resident keys; our cross-device step-up targets the case of a cloud agent acting for a human on a different device. ID-JAG~\cite{idjag} and MCP's enterprise-managed authorization~\cite{mcpauth} give IdPs control over which applications an agent may reach; our design supplies them with stronger evidence rather than replacing them. The FIDO Alliance has formed working groups on agentic authentication and delegation for commerce~\cite{fido2026}, and NIST's NCCoE has begun a project on software and AI agent identity~\cite{nccoe}.

\paragraph{Products.} Enterprise offerings register agents as directory identities with sponsors and conditional access~\cite{entra}, issue certificate-backed agent credentials with human owners and policies~\cite{digicert}, or authorize each tool call at an agent gateway with identity-correlated audit logs~\cite{duo}. None that we are aware of binds an agent's key to attested hardware and to a passkey-signed mandate within one credential.

\paragraph{Research.} South et al.~\cite{south2025} propose authenticated, scoped delegation from humans to AI agents by extending OAuth and OpenID Connect; our work shares the delegation model and adds hardware binding, runtime-enforced origin binding and attestation. Personhood credentials~\cite{adler2024} address the complementary problem of proving that a human, not an AI, is present.

\section{Limitations and Future Work}
\label{sec:limits}

\paragraph{No implementation or measurements yet.} This paper presents a design and an informal analysis. We have not built a full Agent Authenticator, so we report no performance, usability or deployment measurements. A reference implementation on Linux with TPM 2.0, on macOS with the Secure Enclave, and in one confidential-computing environment is the immediate next step.

\paragraph{Formal verification.} The analysis in Section~\ref{sec:analysis} is informal. We plan to model registration, authentication, step-up and sub-delegation in a symbolic protocol verifier such as Tamarin and to state and prove origin binding, approval non-replay and monotonic delegation as formal properties.

\paragraph{Local process identity.} Hardware protects keys from being copied, but most desktop operating systems cannot attest which process requested a signature. The AA's process-lineage check is a software control, and an administrator-level adversary defeats it. Stronger local isolation, for example running agents in lightweight virtual machines with their own attestation, deserves study.

\paragraph{Privacy.} Hardware attestation is linkable across relying parties. Sending only normalized levels mitigates this partially; unlinkable presentation using anonymous credentials or Direct Anonymous Attestation is future work.

\paragraph{Human factors.} Step-up approvals invite approval fatigue. The design limits approvals to actions listed in the mandate and renders exactly what is being signed, but whether people review such prompts carefully needs user studies.

\paragraph{Trust roots and adoption.} Assurance ultimately depends on hardware and cloud vendors' attestation roots. Adoption depends on IdPs and services accepting agent assertions; the bindings in Section~\ref{sec:deploy} are designed to minimize that cost, and we intend to pursue standardization of the protocol.

\section{Conclusion}

Agents need credentials, and today they get the worst kind: ambient, exportable, broadly scoped bearer secrets protected by controls that the agent itself can be talked out of. The Agent Authenticator transfers the properties that made passkeys effective for people, hardware-held keys, per-service scoping and an origin chosen by trusted software rather than by the party being protected, to autonomous agents, and adds what agents uniquely need: human-signed scoped mandates, per-action human approval that works across devices, normalized environment attestation, and records that bind every action to the credential and the person that authorized it. Because it deploys as a profile over MCP, OAuth, ID-JAG, git and cloud identity, it can be adopted incrementally. We hope this design provides a concrete starting point for standardizing how agents authenticate.

\section*{Acknowledgments and Use of Generative AI}

The author used Claude (Anthropic), a large language model, to assist with literature search, drafting and editing text, preparing figures, and producing the \LaTeX{} source. The author reviewed and edited all content, checked every reference against its original source, and takes full responsibility for the paper. The language model is not an author of this work.

\bibliographystyle{unsrt}
\bibliography{refs}

\end{document}